\documentclass[a4paper,11pt]{article}
\usepackage{amssymb,amsmath,bm}
\usepackage{graphics,graphicx}
\usepackage{array,booktabs}
\usepackage{authblk}
\usepackage{slashed}
\usepackage{subfig}
\usepackage{cite}
\usepackage[margin=2.5cm]{geometry}
\usepackage[table,dvipsnames]{xcolor} 
\definecolor{Dblue}{rgb}{0.1, 0.1, 0.8}
\definecolor{Lblue}{rgb}{0.22,0.51,0.9}
\definecolor{green}{RGB}{53,170,102}
\definecolor{Green}{rgb}{0.0, 0.5, 0.0}
\definecolor{Bred}{rgb}{0.8, 0.25, 0.33}
\usepackage{tikz}
\usepackage{multirow}
\usepackage{tikz-feynman}
\usepackage[linktocpage,
colorlinks = true,
linkcolor = Bred,
urlcolor  = Dblue,
citecolor = green,
anchorcolor = blue]{hyperref}
\usepackage{epsfig}
\numberwithin{equation}{section}
\usepackage{multirow}
\usepackage{cite}
\usepackage{xcolor}
\usepackage[margin=2.5cm]{geometry}
\usepackage{orcidlink}
\usepackage[utf8]{inputenc}
\usepackage{orcidlink}

\DeclareUnicodeCharacter{2212}{-}

\numberwithin{equation}{section}

\usepackage{epsfig}

\title{\textbf{Limitations of Freeze-in WIMP Dark Matter from Supercooled Phase Transitions}}

\author[a]{Seyed Yaser Ayazi\thanks{\href{mailto:syaser.ayazi@semnan.ac.ir}{syaser.ayazi@semnan.ac.ir}}}
\author[a]{ Mojtaba Hosseini\thanks{\href{mailto:mojtaba\textunderscore hosseini@semnan.ac.ir}{mojtaba\textunderscore hosseini@semnan.ac.ir}}}

\affil[a]{Department of Physics, Semnan University, P.O. Box 35131-19111, Semnan, Iran}
\date{\today}

\begin{document}

\baselineskip 0.6 cm
\maketitle

\begin{abstract}
	We revisit the possibility of producing Weakly Interacting Massive Particle (WIMP) dark matter via a freeze-in mechanism triggered by a supercooled first-order phase transition (FOPT) in the early universe. Unlike traditional freeze-out and FIMP scenarios, this mechanism relies on a rapid entropy injection that dilutes the preexisting dark matter abundance and prevents re-equilibration due to a sudden mass increase. In this study, we systematically examine a variety of single-component dark matter models—including vector, fermionic, and scalar-mediated candidates—to assess whether they can satisfy the key cosmological condition \( T_2 \gg T_1 \), required for successful WIMP freeze-in after FOPT. Contrary to earlier results, our revised analysis finds that \textbf{none} of the models fulfill this condition across viable parameter spaces. We confirm, however, that the scalar dark matter model analyzed in Ref.~\cite{Wong:2023qon} is the \textbf{only known viable single-component model} that fulfills \( T_2 \gg T_1 \) and enables WIMP freeze-in via this mechanism. These findings place important constraints on model-building efforts and suggest that successful freeze-in after FOPT may require \textit{multi-component or more complex dark sectors} beyond the scope of minimal models.

\end{abstract}



\section{Introduction} \label{sec1}
The nature of DM remains one of the most profound mysteries in modern physics\cite{Bertone:2016nfn,Cirelli:2024ssz}. Despite overwhelming astrophysical and cosmological evidence for its existence, the particle identity of DM continues to elude discovery. The Standard Model of elementary particles(SM), despite its enormous success in describing nature in its most fundamental form, does not contain any candidates for DM. Therefore, the presence of models beyond the SM is essential to describe DM\cite{Zurek:2008qg,Profumo:2009tb,Aoki:2012ub,Biswas:2013nn,Gu:2013iy,Aoki:2013gzs,Kajiyama:2013rla,Bian:2013wna,Bhattacharya:2013hva,Geng:2013nda,Esch:2014jpa,Dienes:2014via,Bian:2014cja,Geng:2014dea,DiFranzo:2016uzc,Aoki:2016glu,DuttaBanik:2016jzv,Pandey:2017quk,Borah:2017xgm,Herrero-Garcia:2017vrl,Ahmed:2017dbb,PeymanZakeri:2018zaa,Aoki:2018gjf,Chakraborti:2018lso,Bernal:2018aon,Poulin:2018kap,Herrero-Garcia:2018qnz,YaserAyazi:2018lrv,Elahi:2019jeo,Borah:2019epq,Borah:2019aeq,Bhattacharya:2019fgs,Biswas:2019ygr,Nanda:2019nqy,Yaguna:2019cvp,Belanger:2020hyh,VanDong:2020bkg,Khalil:2020syr,DuttaBanik:2020jrj,Hernandez-Sanchez:2020aop,Chakrabarty:2021kmr,Yaguna:2021vhb,DiazSaez:2021pmg,DiazSaez:2021pfw,Mohamadnejad:2021tke,Yaguna:2021rds,Ho:2022erb,Kim:2022sfg,Das:2022oyx,Belanger:2022esk,Hosseini:2023qwu,YaserAyazi:2024hpj,YaserAyazi:2022tbn,YaserAyazi:2019caf,YaserAyazi:2024dxk,Abkenar:2024ket}. Among the many theoretical paradigms proposed to explain DM, WIMPs have been extensively studied, primarily through the well-established freeze-out mechanism\cite{Feng:2022rxt,Roszkowski:2017nbc,Arcadi:2017kky}. In this scenario, WIMPs achieve thermal equilibrium in the early universe and decouple as the universe expands and cools. However, traditional WIMP models face challenges from stringent experimental constraints, motivating the exploration of alternative production mechanisms.

The freeze-in mechanism offers a compelling alternative, wherein DM particles are never in thermal equilibrium but are produced gradually from the interactions of particles in the thermal bath\cite{Hall:2009bx}. Freeze-in has historically been associated with FIMPs, which interact so weakly that their production rate is suppressed, resulting in a relic density that matches observations.

The role of FOPT and its impact on DM phenomenology is very important\cite{Hambye:2018qjv,Baker:2019ndr,Chway:2019kft,Cohen:2008nb,Baker:2017zwx,Bian:2018mkl,Bian:2018bxr,Baker:2016xzo,Kobakhidze:2017ini,Baker:2018vos,DiBari:2020bvn,Kobakhidze:2019tts,Falkowski:2012fb,Baldes:2020kam,Azatov:2021ifm,Baldes:2022oev,Baldes:2021aph,Roy:2022gop,Krylov:2013qe,Huang:2017kzu,Bai:2018vik,Bai:2018dxf,Atreya:2014sca,Hong:2020est,Baker:2021nyl,Kawana:2021tde,Liu:2021svg,Baker:2021sno,Hashino:2021qoq,Huang:2022him,Bai:2022kxq,Kawana:2022lba,He:2022amv,Baldes:2023cih,Elor:2021swj}.
 Recently, however, a novel variation of the freeze-in mechanism has been proposed for WIMPs, where weak or moderate interaction strengths suffice due to the influence of a supercooled FOPT in the early universe\cite{Wong:2023qon}(Freeze-in of WIMP). This scenario opens a new path for DM model building, enabling WIMPs to remain a viable candidate while resolving challenges associated with traditional freeze-out. It provides a robust framework for exploring physics beyond the SM, including connections to FOPT and GW signals. This new mechanism suggests WIMPs can also freeze-in under certain conditions:
\begin{itemize}
	\item A \textit{supercooled first-order phase transition} dilutes preexisting DM density to near zero.
	\item The phase transition increases the mass of DM significantly, leading to a large mass-to-temperature ratio. This prevents WIMPs from re-equilibrating with the thermal bath.
\end{itemize}

Therefore, in the study of DM production in the early universe, two main mechanisms are commonly considered: the traditional WIMP freeze-out
and FIMP freeze-in. These mechanisms differ in their interaction strengths, thermal equilibrium conditions, and the role of phase transitions. Freeze-out describes the decoupling of once-equilibrated DM particles, leading to their relic abundance. Traditional freeze-in applies to feebly interacting particles that never reached equilibrium but were gradually produced over cosmic history. The WIMP freeze-in provides a new connection between FOPT and DM, opening up a third possibility for realizing DM besides WIMPs and FIMPs. It can be tested by combining the WIMP and GW experiments.

This work explores the WIMP freeze-in production of various DM candidates facilitated by an FOPT. The FOPT provides two essential effects: the injection of entropy that dilutes preexisting DM and a rapid mass increase for DM particles, preventing thermal equilibration. While the scalar WIMP candidate is a natural choice for demonstrating this mechanism\cite{Wong:2023qon}, it can be extended to several other DM candidates.

The paper is organised as follows. In section.~\ref{Models}, we extend SM by several different models and DM candidates. In section.~\ref{FOPT}, we examine the conditions of the WIMP freeze-in scenario and present the numerical analyses of the models in light of supercooled FOPT scenario. Finally, the conclusions present in section.~\ref{Conclusion}.

\section{The models}\label{Models}
In this section, we present different models with different DM candidates, for which we will examine the conditions of the Freeze-in of WIMP in the next section.
These models are presented in Table \ref{dmcandidates}.

\subsection{Vector DM with $U(1)$ symmetry}\label{2.1 model}
The Lagrangian of this model is as follows\cite{YaserAyazi:2024hpj,YaserAyazi:2022tbn}:
\begin{equation}
 {\cal L} ={\cal L}_{SM}- \frac{1}{4} V_{\mu \nu} V^{\mu \nu}+ (D'_{\mu} S)^{*} (D'^{\mu} S) - V(H,S)  , \label{3-1}
\end{equation}
where $ {\cal L} _{SM} $ is the SM Lagrangian without the Higgs potential term and the covariant derivative and field strength of $V_{\mu}$ are given as:
\begin{align}
& D'_{\mu} S= (\partial_{\mu} + i g_v V_{\mu}) S,\nonumber \\
& V_{\mu \nu}= \partial_{\mu} V_{\nu} - \partial_{\nu} V_{\mu}.\nonumber \end{align}
The potential which is renormalizable and invariant
under gauge and $ Z_{2} $ symmetry is:
\begin{equation}
V(H,S) = -\mu_{H}^2 H^{\dagger}H-\mu_{S}^2 S^*S+\lambda_{H} (H^{\dagger}H)^{2} + \lambda_{S} (S^*S)^{2} +  \lambda_{S H} (S^*S) (H^{\dagger}H). \label{3-2}
\end{equation}
Due to the imposed $Z_2$ symmetry under which only the dark gauge
boson $V_{\mu}$  is odd, the vector field $V_{\mu}$ can be considered as a DM candidate. The $S$ field is the intermediate between the dark sector and the SM sector. After the electroweak symmetry breaking, the parameters of the model are given by:	
\begin{align}
& \nu _2=\frac{M_V}{g_v} \nonumber,~~~~~~~~~~ \sin\alpha=\frac{\nu_1}{\sqrt{\nu _1^2+\nu_2^2}} \\
& \lambda _H=\frac{\cos ^2\alpha M_{H_1}^2+\sin ^2\alpha  M_{H_2}^2}{2 \nu _1^2}  \nonumber \\
& \lambda _S=\frac{\sin ^2\alpha M_{H_1}^2+\cos ^2\alpha  M_{H_2}^2}{2 \nu _2^2}  \nonumber \\
&  \lambda _{SH}=\frac{ \left(M_{H_2}^2-M_{H_1}^2\right) \sin \alpha  \cos \alpha}{\nu _1 \nu _2} \label{3-3}
\end{align}
Where the model includes three independent parameters $M_V $, $ M_{H_2} $ and $ g_v $.
	
\subsection{Scale Invariant Vector DM with $U(1)$ symmetry}\label{2.2 model}
In this model, there is no mass term at the tree level, and electroweak symmetry breaking occurs at the one-loop level via the Gildener-
Weinberg mechanism\cite{Gildener:1976ih}. The Lagrangian of this model is as follows\cite{YaserAyazi:2019caf,Hosseini:2024ahe}:
\begin{equation}
 {\cal L} ={\cal L}_{SM} + (D'_{\mu} S)^{*} (D'^{\mu} S) - V(H,S) - \frac{1}{4} V_{\mu \nu} V^{\mu \nu} , \label{3-4}
\end{equation}
where $ {\cal L} _{SM} $ is the SM Lagrangian without the Higgs potential term and
\begin{align}
& D'_{\mu} S= (\partial_{\mu} + i g_vV_{\mu}) S,\nonumber \\
& V_{\mu \nu}= \partial_{\mu} V_{\nu} - \partial_{\nu} V_{\mu}.\nonumber \end{align}
The most general scale-invariant potential which is renormalizable and invariant
under gauge and $ Z_{2} $ symmetry is:
\begin{equation}
V(H,S) = \frac{\lambda_{H}}{6} (H^{\dagger}H)^{2} +\frac{\lambda_{S}}{6}  (S^*S)^{2} + 2 \lambda_{S H} (S^*S) (H^{\dagger}H). \label{3-5}
\end{equation}
Similar to the previous model, the vector field $ V_{\mu} $ is a candidate for DM and $S$ is the intermediate between the dark sector and the SM.
After the symmetry breaking, we have the following constraints:
\begin{align}
& \nu_{2} =  \frac{M_{V}}{g_v} , &\nonumber
& sin \alpha =  \frac{\nu_{1}}{\sqrt{\nu_{1}^{2}+\nu_{2}^{2}}} \nonumber \\
& \lambda_{H} =  \frac{3 M_{H_{1}}^{2}}{ \nu_{1}^{2}} cos^{2} \alpha  \nonumber&
& \lambda_{S} =  \frac{3 M_{H_{1}}^{2}}{ \nu_{2}^{2}} sin^{2} \alpha  &\nonumber\\
& \lambda_{S H} =  - \frac{ M_{H_{1}}^{2}}{2 \nu_{1} \nu_{2} } sin \alpha \, cos \alpha , \label{3-6}
\end{align}
where $M_V$ is the mass of vector DM after symmetry breaking. Regarding one-loop effect, the dark Higgs mass is given by\cite{YaserAyazi:2019caf,Gildener:1976ih}
\begin{equation}
M_{H_{2}}^{2} = \frac{1}{8 \pi^{2} \nu^{2}} \left( M_{H_{1}}^{4} + 6  M_{W}^{4} + 3  M_{Z}^{4} + 3  M_{V}^{4} - 12 M_{t}^{4} \right)\label{3-7},
\end{equation}
where $ M_{W,Z,t}$ are the masses of W, Z gauge bosons, and top quark, respectively. Constraints (\ref{3-6}) severely restrict free parameters of the model up to two independent parameters of $M_V$ and $g_v$.

\subsection{Fermionic DM with $U(1)$ symmetry}\label{2.3 model}
This model is also scale invariant and symmetry breaking occurs at the one-loop level. The model contains three new fields: a vector field $V_{\mu}$, a Dirac fermion field $\psi$ that can play the role of DM and a complex scalar $S$, mediates between SM and the dark sector. In the model $V_{\mu}$, $\psi$ and $S$ are charged under a new dark $U(1)_D$ gauge group and all of these fields are singlet under SM gauge groups. The $U(1)_D$ charge of the new particles are given in Table \ref{Table1}.
The Lagrangian of this model is as follows\cite{Hosseini:2023qwu}:
\begin{align}
{\cal L} ={\cal L}_{SM}+ i\bar\psi_L  \gamma^{\mu}D_{\mu}\psi_L+ i \bar\psi_R \gamma^{\mu}D_{\mu}\psi_R
-g_s \bar\psi_L \psi_R S+h.c.-\frac{1}{4} V_{\mu \nu} V^{\mu \nu}+ (D_{\mu} S)^{*} (D^{\mu} S)- V(H,S).
\label{eq:lagrangian}
\end{align}
where $ {\cal L} _{SM} $ is the SM Lagrangian without the Higgs potential term, The covariant derivative is
\begin{align}
& D_{\mu}= (\partial_{\mu} + i Q g_{v} V_{\mu}), \quad \text{and}\nonumber \\
& V_{\mu \nu}= \partial_{\mu} V_{\nu} - \partial_{\nu} V_{\mu},
\end{align}
where $Q$ is the charge of the new particles under $U(1)$ symmetry.
The most general scale-invariant potential $ V(H,S) $ that is renormalizable and invariant
under gauge symmetry is
\begin{equation}
V(H,S) = \frac{1}{6} \lambda_{H} (H^{\dagger}H)^{2} + \frac{1}{6} \lambda_{S} (S^{*}S)^{2} + 2 \lambda_{S H} (S^{*}S) (H^{\dagger}H).
\end{equation}
After the symmetry breaking, we have the following constraints:
\begin{align}
& \nu_{2} =  \frac{M_{V}}{g_v} , &\nonumber
& sin \alpha =  \frac{\nu_{1}}{\sqrt{\nu_{1}^{2}+\nu_{2}^{2}}} \nonumber \\
&M_{\psi}= \frac{g_sM_{V}}{\sqrt{2}g_v}&\nonumber
& \lambda_{H} =  \frac{3 M_{H_{1}}^{2}}{ \nu_{1}^{2}} cos^{2} \alpha  \nonumber  \\
& \lambda_{S} =  \frac{3 M_{H_{1}}^{2}}{ \nu_{2}^{2}} sin^{2} \alpha  &\nonumber
& \lambda_{S H} =  - \frac{ M_{H_{1}}^{2}}{2 \nu_{1} \nu_{2} } sin \alpha \, cos \alpha , \\ \label{2-11}
\end{align}
where $M_{\psi}$ and $M_{V}$ are the masses of vector and fermion fields after symmetry breaking. Meanwhile, since we do not assume $M_V<2M_{\psi}$, the only candidate for DM is $\psi$. Regarding one-loop effect, the dark Higgs mass is given by\cite{Gildener:1976ih}:
\begin{equation}
M_{H_{2}}^{2} = \frac{1}{8 \pi^{2} \nu^{2}} \left( M_{H_{1}}^{4} + 6  M_{W}^{4} + 3  M_{Z}^{4} + 3  M_{V}^{4} - 12 M_{t}^{4} -4M_{\psi}^{4}  \right) .
\end{equation}
where $ M_{W,Z,t} $ are the masses of W, Z gauge bosons, and top quark, respectively. In the equations above, $H_1$ is the Higgs boson and $ M_{H_{1}} = 125 $ GeV.
According to  (\ref{2-11}), the independent parameters of the model are $ M_{V} $, $M_{\psi}$ and the coupling $ g_v $ .

\begin{table}
\begin{center}
\begin{tabular}{| l | l | l | l |}
\hline
Field&$S$&$\psi_L$&$\psi_R$\\ \hline
$U(1)_D$ charge&1&$\frac{1}{2}$&$-\frac{1}{2}$\\ \hline
\end{tabular}
\end{center}
\caption{\label{Table1}The charges of the dark sector particles under the new $U(1)_D$ symmetry.}
\end{table}

\subsection{Fermionic DM with scalar mediator vs. pseudoscalar mediator}\label{2.4 model}
In this model, the SM is extended with two new fields: a Dirac fermion $\psi$ which plays the role of the DM candidate and a (pseudo)scalar field $S$ that interacts with the DM through a Yukawa term. The Lagrangian of the model is as follows\cite{Ghorbani:2017jls}:
\begin{align}
{\cal L} ={\cal L}_{SM}+{\cal L}_{dark}(\psi,S)+{\cal L}_{S}(S)+{\cal L}_{int}(H,S,\psi),
\end{align}
where ${\cal L}_{SM}$ stands for the SM Lagrangian. The Lagrangian of fermionic DM, ${\cal L}_{dark}(\psi,S)$, is :
\begin{align}
{\cal L}_{dark}(\psi,S)=\bar\psi(i\gamma_{\mu}\partial^{\mu}-M_{\psi})\psi.
\end{align}
${\cal L}_{S}$ is as follows:
\begin{align}
{\cal L}_{S}=\frac{1}{2}(\partial^{\mu} S)^2 -\frac{1}{2}\mu_{S}^2S^2 -\frac{1}{4}\lambda_{S}S^4,
\end{align}
and ${\cal L}_{int}$ is the (pseudo)scalar interaction with the dark and the SM sectors. When $S$ is a pseudoscalar then,
\begin{align}
{\cal L}_{int}(H,S,\psi)=g_d S \bar\psi \gamma^5 \psi + \frac{1}{2}\lambda_{SH}S^2 H^{\dagger}H,
\end{align}
and when $S$ represents the scalar,
\begin{align}
{\cal L}_{int}(H,S,\psi)=g_d S \bar\psi \psi + \frac{1}{2}\lambda_{SH}S^2 H^{\dagger}H.
\end{align}
The Higgs potential in the SM sector reads:
\begin{align}
V_H=-\mu_{H}^2H^{\dagger}H-\lambda_H (H^{\dagger}H)^2.
\end{align}
The tree-level potential which is given by the
substitution $H=(0 ~~h\equiv\langle h_1 \rangle)^{\dagger}$ and $S\equiv\langle h_2\rangle$  reads,
\begin{align}
V_0(h_1 , h_2)=-\frac{1}{2}\mu_{H}^2h_1^2 - \frac{1}{2}\mu_{S}^2h_2^2+\frac{1}{4}\lambda_H h_1^4 +\frac{1}{4}\lambda_S h_2^4+\frac{1}{2}\lambda_{SH}h_1^2 h_2^2. \label{2-19}
\end{align}
 After the symmetry breaking both Higgs particle and the scalar field undergo non-zero $vev$. In fact, by substituting $h_1 \longrightarrow \nu_1 + h_1$ and $h_2 \longrightarrow \nu_2 + h_2$, the fields $h_1$ and $h_2$ mix with each other and they can be rewritten by the mass eigenstates $H_1$ and $H_2$ as
 \begin{align}
& H_1 = h_1 \cos\theta +h_2 \sin\theta , \nonumber\\
& H_2= -h_1 \sin\theta + h_2 \cos\theta.
\end{align}
In the above equations, $\theta$ is the mixing angle. In the following we assume that $H_1$ is the eigenstate of the SM Higgs with $M_{H_1} = 125$ GeV ($\nu_1=246$ GeV) and $H_2$ corresponds to the eigenstate of the singlet scalar. After symmetry breaking, we have
\begin{align}
& \lambda _H=\frac{ M_{H_2}^2\sin ^2\theta+M_{H_1}^2\cos ^2\theta}{2 \nu _1^2}  \nonumber \\
& \lambda _S=\frac{ M_{H_2}^2\cos ^2\theta+M_{H_1}^2\sin ^2\theta}{2 \nu _2^2}  \nonumber \\
&  \lambda _{SH}=\frac{ \left(M_{H_2}^2-M_{H_1}^2\right) \sin 2\theta}{2\nu _1 \nu _2}
\end{align}
According to the above equations, the independent parameters of the model can be chosen as $M_{H_2}$,$M_\psi$,$ \theta$, $g_d$ and $\nu_2$.
Note that all we have said in this section is true for both scalar and pseudoscalar mediators.

\begin{table}[h]
	\centering
	\renewcommand{\arraystretch}{1.3} 
	\arrayrulecolor{black} 
	\rowcolors{2}{cyan!15}{white} 
	\begin{tabular}{|c|c|c|}
		\hline
		\rowcolor{gray!30} 
		\textbf{Models} & \textbf{Symmetry} & \textbf{Key Parameters} \\
		\hline
		Vector DM with \( U(1) \) symmetry & \( U(1) \) & \( M_V, M_{H_2}, g_v \) \\
		Scale Invariant Vector DM with \( U(1) \) symmetry & \( U(1) \) & \( M_V, g_v \) \\
		Fermionic DM with \( U(1) \) symmetry & \( U(1) \) & \( M_V, M_\psi, g_v \) \\
		Fermionic DM with Scalar Mediator & \( Z_2 \) & \( M_{H_2}, M_\psi, \theta, g_d, \nu_2 \) \\
		Fermionic DM with Pseudoscalar Mediator & \( Z_2 \) & \( M_{H_2}, M_\psi, \theta, g_d, \nu_2 \) \\
		\hline
	\end{tabular}
	\caption{Summary of DM models, their symmetries, and key parameters.}
	\label{dmcandidates}
\end{table}

\section{Freeze-in of WIMP}\label{FOPT}
Freeze-in of WIMP is a novel scenario for producing DM that is the boundary between freeze-out and freeze-in mechanisms\cite{Wong:2023qon}.
In this mechanism, in the conventional thermal history of the Universe, DM particles inevitably thermalize and freeze-out. Freeze-in of WIMPs can happen if the Universe experiences a supercooled FOPT. A supercooled FOPT releases a huge amount of entropy, which dilutes the preexisting DM density to a negligible level.
The WIMPs gain mass from the FOPT and never return to equilibrium due to their large mass-to-temperature ratio. After the FOPT, DM will be accumulatively produced via the process of $SM~SM \rightarrow DM~DM$, which is a typical freeze-in scenario, but it applies to weak or moderate couplings.
The main point of this introduced mechanism is that the coupling between the dark side and the SM is larger than the conventional freeze-in case.

In the early universe, DM($X$) was massless and $z\equiv (M_X/T)=0$. With the start of supercooling at temperature $T_2$, the DM mass undergoes a sudden change to $M_X \gg T_2$, where $T_2$ denotes the temperature after the FOPT. We assume a supercooled FOPT such that $T_2 \gg T_1$.
 $T_1 $ is the temperature at which the FOPT begins. This leads to an enormous increase in entropy density by a factor of $(T_2/T_1)^3$ and dilutes the preexisting DM density to a negligible level. Consequently, the evolution of the density of DM begins at $z_2 = M_X/T_2 \gg 1$, with
an initial condition $Y_X(z_2) \approx (T_1/T_2)^3Y_1 \sim 0$. From here on, the production of DM particles begins with the freeze-in mechanism and with a negligible initial amount.

With the above interpretations, the two basic conditions for establishing the presented scenario are as follows:
\begin{itemize}
 	\item \textbf{Reheating temperature $T_2$ must be greater than the critical temperature $T_1$($T_2 \gg T_1$)}:

 In this scenario, $T_2$(reheating temperature) must always be greater than $T_1$. $T_\Lambda$ is the temperature at which supercooling(starting a phase of thermal inflation) begins and when reheating is prompt ($\Gamma\gg H$), we have\cite{Marfatia:2020bcs,Hambye:2018qjv}:
 \begin{align}\label{3.1equation}
T_2= T_\Lambda = (\frac{30 V_\Lambda}{\pi^2 g_{*}})^{\frac{1}{4}}.
\end{align}
In the above equation, $g_{*} \approx 106.75$ and for all models presented in the paper, $\Gamma\gg H$, where $\Gamma$ is the decay width of scalars and $H$ is the Hubble constant during the FOPT. The $V_\Lambda$ is equal to the energy difference between the true and false vacuums during the phase transition. The $V_\Lambda$ quantity is also defined as follows\cite{Hambye:2018qjv}:
\begin{align}\label{3.2equation}
V_\Lambda= \frac{\beta_{\lambda_S}v_2^4}{16},
\end{align}
where $v_2$ is the vacuum expectation value of the scalar mediator $S$ and $\beta_{\lambda_S}$ is the one-loop beta function for coupling of scalar mediator $\lambda_S$.
Of course, equation \ref{3.2equation} only applies to scale invariant models, i.e. models \ref{2.2 model} and \ref{2.3 model}. In the following, we examine this condition for all the presented models.
\begin{enumerate}
	\item The Model \ref{2.1 model} :

This medel includes three independent parameters $M_V $, $ M_{H_2} $ and $ g_v $. For this model, all possible phase transition states have been investigated in reference \cite{Hashino:2018zsi}. In this case, there is no phase transition, either one-step or two-step, that can be observed in consistent with the DM constraints, including the correct relic density of DM\cite{Planck:2018vyg}. Therefore, this model cannot be included in the mechanism proposed for the production of DM in this paper.

\begin{figure}
	\begin{center}
		\centerline{\hspace{0cm}\epsfig{figure=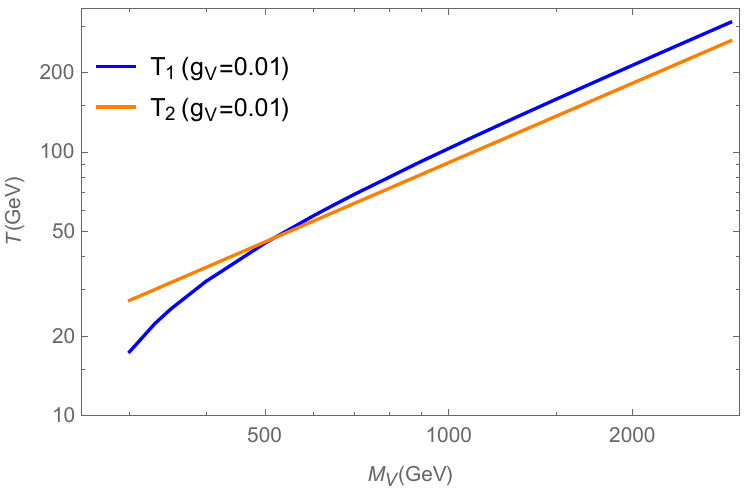,width=8cm}\hspace{0.5cm}\hspace{0cm}\epsfig{figure=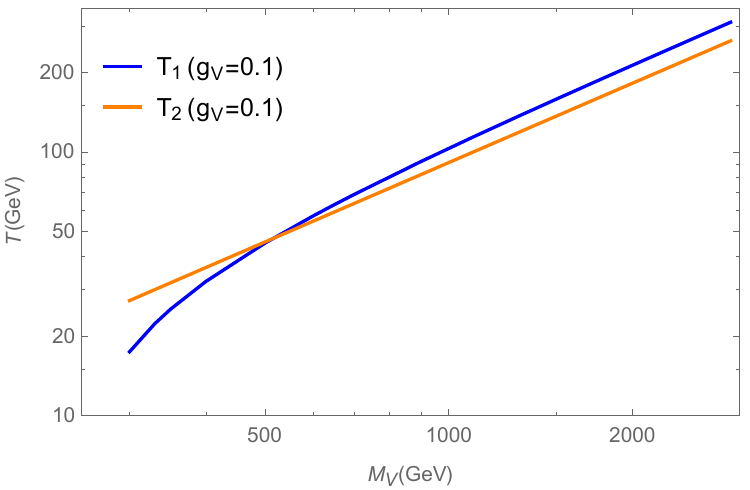,width=8cm}}
		\centerline{\vspace{0.2cm}\hspace{1.2cm}(a)\hspace{8cm}(b)}
		\centerline{\hspace{0cm}\epsfig{figure=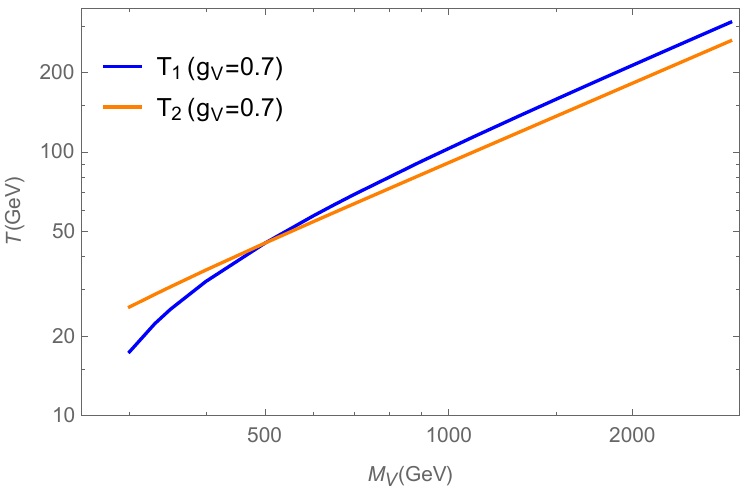,width=8cm}\hspace{0.5cm}\hspace{0cm}\epsfig{figure=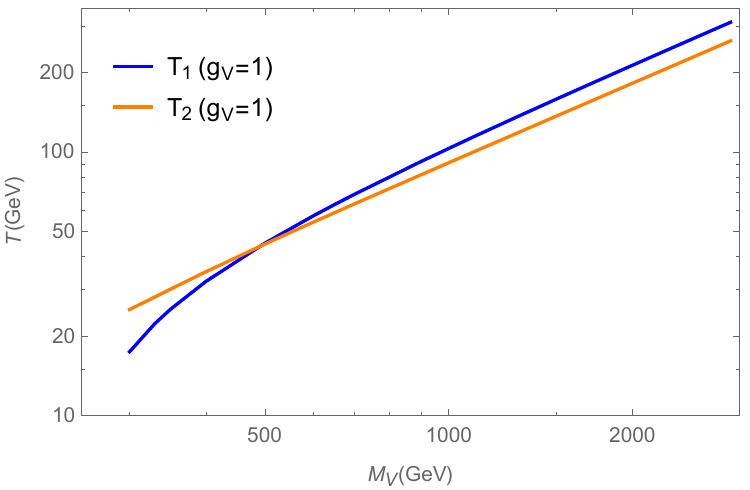,width=8cm}}	
		\centerline{\vspace{0.2cm}\hspace{0.8cm}(c)\hspace{8cm}(d)}
		\caption{Temperature variations for the model \ref{2.2 model} with respect to DM mass for different couplings.} \label{vector invariant}
	\end{center}
\end{figure}

\begin{figure}
	\begin{center}
		\centerline{\hspace{0cm}\epsfig{figure=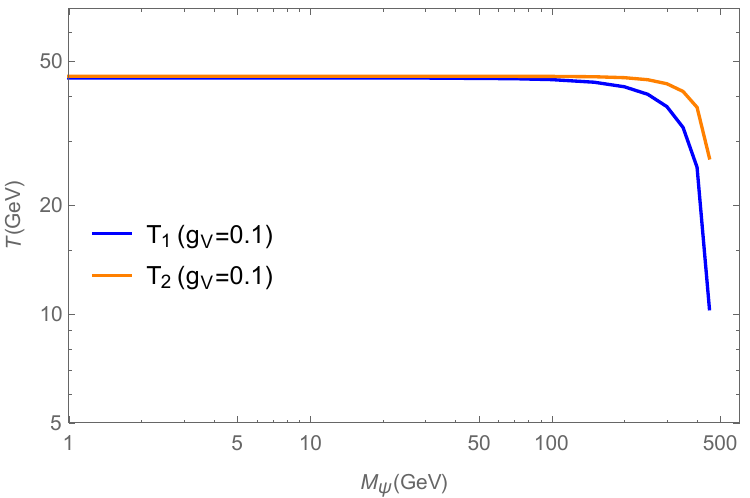,width=8cm}\hspace{0.5cm}\hspace{0cm}\epsfig{figure=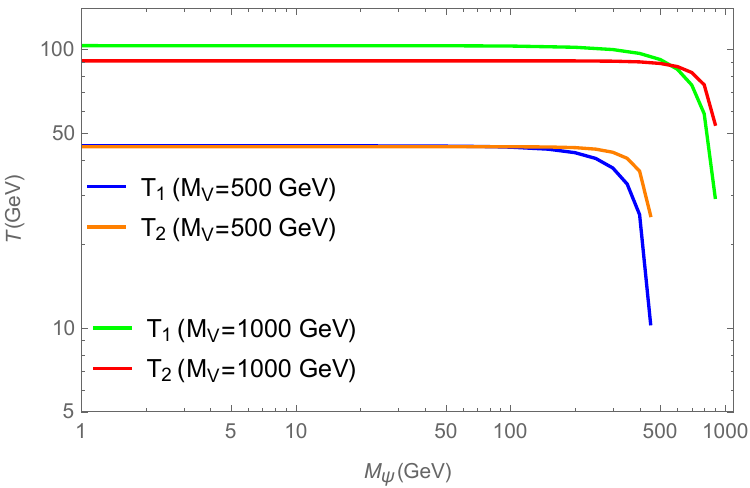,width=8cm}}
		\centerline{\vspace{0.2cm}\hspace{1.2cm}(a)\hspace{8cm}(b)}
		\caption{Temperature variations for the model \ref{2.3 model} with respect to DM mass. In (a) $M_V=500$ GeV and in (b) $g_v=1$ is considered. We know that in the model must always $M_\psi < M_V/2$ so that the only DM particle in the model is the fermion $\psi$.} \label{fermion invariant}
	\end{center}
\end{figure}

\item The Model \ref{2.2 model} :

This medel includes two independent parameters $M_V $ and $ g_v $. According to equation \ref{3-7}, $M_V$ must always be greater than 240 GeV.
Figure \ref{vector invariant} shows the behavior of critical temperature($T_1$) and reheating temperature($T_2$) for different couplings and masses.
Details on the calculation of $T_1$ are given in the Appendix. As can be seen, only for a very small portion of the parameter space(240 GeV$ < M_V<$ 500 GeV), $T_2$ is larger than $T_1$.
But to establish the mechanism of Freeze-in of WIMP, we must assume that $T_2 \gg T_1$, which is not the case here, and both temperatures are very close and $T_2/T_1\sim 1$. Therefore, this model cannot be included in the framework of the presented mechanism.

\item The Model \ref{2.3 model} :

This medel includes three independent parameters $M_V $, $M_{\psi}$ and $ g_v $. First, it is important to note that in the model, for our only candidate for DM to be a fermionic particle($\psi$), we must have $M_\psi < M_V /2$. In Figure \ref{fermion invariant}(a), temperature changes are plotted against the mass of DM for constant $g_v$ and $M_V$. As can be seen, $T_2 \sim T_1$ and $T_2/T_1\sim 1$. But from $M_\psi > 250$ GeV, $T_2$ will be a relatively large distance from $T_1$, which is not the case in the model scenario where only the fermionic particle is a candidate for DM($M_\psi < M_V /2 = 250$ GeV).
In Figure \ref{fermion invariant}(b), temperature changes are plotted for different $M_V$. As can be seen, as $M_V$ increases, $T_2$ becomes smaller than $T_1$.
But for smaller $M_V$, $T_2$ becomes a relatively significant distance from $T_1$ at a point where the single-component DM condition(only $\psi$ as a DM candidate) in the model is violated(because of $M_\psi > M_V/2$). Therefore, this model like the previous two models cannot be included in the framework of the presented mechanism.
\item The Model \ref{2.4 model} :

Since the model \ref{2.4 model} with scalar mediator is very limited and it has been publicly ruled out by direct discovery experiments\cite{Ghorbani:2017jls}, we examine Model \ref{2.4 model} with pseudoscalar mediator.
By choosing the mediator in the model \ref{2.4 model} to be a pseudoscalar\cite{Ghorbani:2017jls}, the elastic scattering cross section for DM-nucleon is velocity suppressed and so the theory easily evades the direct detection bounds from XENONnT and LZ\cite{XENON:2023cxc,LZ:2022lsv}.
There are five independent parameters $M_{H_2}$,$M_\psi$,$ \theta$, $g_d$ and $\nu_2$ in this model. An overall result of the Higgs signal strength measured by ATLAS and CMS will restrict the Higgs mixing angle to values smaller than $\sin \theta \leq 0.12$ $(\theta \leq 6.9^\circ)$\cite{ATLAS:2016neq}. Therefore, in the rest of the paper we consider the mixing angle$(\theta=1^\circ)$. Also, the mass of DM($M_\psi$) will not have any effect on determining the temperatures $T_1$ and $T_2$.

Reference \cite{Ghorbani:2017jls} contains the equations for calculating critical temperature $T_1$. As can be seen, $T_1$ is very small, and by calculating the energy difference between true and false vacuums($V_\Lambda$), we concluded that $T_2$ is of the same order as $T_1$, and this model also does not fit into the  mechanism of freeze in of WIMP. In fact, the perturbation constraint $\lambda_S < 4\pi$ causes a limitation in the range of parameter values and leads to a very small critical temperature, which naturally leads to a small temperature $T_2$ of the same order as $T_1$. Finally, all the models presented in this paper lead to a strong FOPT, but they do not apply to the condition of $T_2\gg T_1$.

A summary of the analysis performed in this section is provided in Table \ref{tab:T1_T2_results}.
\end{enumerate}
\end{itemize}

\begin{table}[h]
	\centering
	\renewcommand{\arraystretch}{1.3} 
	\arrayrulecolor{black} 
	\rowcolors{2}{gray!15}{white} 
	\begin{tabular}{|c|c|}
		\hline
		\rowcolor{cyan!15}
		\textbf{Models} & \textbf{Condition of $T_2 \gg T_1$}  \\
		\hline
		Vector DM with \( U(1) \) symmetry  & Impossible \\
		Scale Invariant Vector DM with \( U(1) \) symmetry  & Impossible \\
		Fermionic DM with \( U(1) \) symmetry & Impossible \\
		Fermionic DM with Scalar Mediator & Impossible \\
	    Fermionic DM with Pseudoscalar Mediator & Impossible \\
		\hline
	\end{tabular}
	\caption{Condition of $T_2 \gg T_1$ for different models}
	\label{tab:T1_T2_results}
\end{table}

\begin{itemize}
 	\item \textbf{Failure to return to equilibrium conditions after FOPT ($T_2< T_{dec}$)}:

 In this scenario, for $T_2>T_{dec}$ , the plasma thermalizes again, and the usual freeze-out mechanism yields
the relic abundance\cite{Marfatia:2020bcs,Hambye:2018qjv}. $T_{dec}\sim M_{DM}/25$ is temperature of freeze-out. So we are looking for points in the parameter space where $T_2<T_{dec}$.
\end{itemize}

\section{Conclusion}\label{Conclusion}

In this study, we reassessed the freeze-in production of Weakly Interacting Massive Particles (WIMPs) facilitated by a supercooled FOPT in the early universe. The mechanism relies on the sudden injection of entropy and rapid mass gain following FOPT, which dilutes preexisting dark matter and inhibits thermal re-equilibration, thereby enabling freeze-in at moderate couplings.

We analyzed several classes of single-component dark matter models, including vector dark matter with and without scale invariance, fermionic dark matter with U(1) symmetry, and fermionic dark matter with scalar or pseudoscalar mediators. All models were tested against the key requirement \( T_2 \gg T_1 \), essential for decoupling dark matter from the thermal bath post-FOPT.

Our revised calculations reveal that \textbf{none of these models satisfy the condition \( T_2 \gg T_1 \)} within viable parameter space. These models thus cannot realize the scenario of  freeze-in of WIMP as originally hoped.

The only known single-component model that does meet this condition is the \textbf{scalar dark matter model previously studied in Ref.~\cite{Wong:2023qon}}. This case remains a unique example where the freeze-in WIMP mechanism can be consistently realized through a supercooled FOPT.

These findings indicate that the minimal freeze-in WIMP framework is highly constrained and may require more complex constructions—such as multi-component dark sectors, richer symmetry structures, or non-standard cosmological histories—to become a broadly viable mechanism. Future work should explore such extended scenarios and their phenomenological signatures, including gravitational wave signals from FOPTs.

\section*{Appendix: Calculations related to $T_1$ and $\beta_{\lambda_S}$}

In the high temperature approximation $M_i^2 /T^2\ll1$ for all $i$ with $M_i$ being the mass of the particle $i$ in the model, the potential of Model \ref{2.2 model} is as follows(of course, we know that the potential of the tree level is zero.)\cite{YaserAyazi:2019caf}:
\begin{equation}\label{appendix 1}
V_{eff}^{1-loop}(H_2 ,T) = b H_{2}^{4}(\ln \frac{H_{2}^{2}}{\nu^{2}}-\frac{1}{2})+cT^2H_{2}^{2},
\end{equation}
where

\begin{align}
& b =  \frac{1}{64 \pi^{2} \nu^{4}} (M_{H_{1}}^{4}+6M_{W}^{4}+3M_{Z}^{4}+3M_{V}^{4}-12M_{t}^{4}) , \nonumber \\
& c = \frac{1}{12 \nu^{2}} (M_{H_{1}}^{2}+6M_{W}^{2}+3M_{Z}^{2}+3M_{V}^{2}+6M_{t}^{2}) . \label{appendix 2}
\end{align}
The phase transition takes place at the critical temperature $T_C$($T_1$) at which the finite
temperature one-loop effective potential (\ref{appendix 1}) has two degenerate minimums at $H_2$ = 0 and
$H_2 = \nu_C$, i.e.,
\begin{align}\label{appendix 3}
&V_{eff}^{1-loop}(0 ,T_C) = V_{eff}^{1-loop}(\nu_C ,T_C)=0, \nonumber \\
& \left. \frac{V_{eff}^{1-loop}(H_2 ,T_C)}{dH_2} \right|_{H_2=\nu_C}=0.
\end{align}
By simultaneously solving the above equations, the critical temperature is obtained as follows:
\begin{equation}\label{critical1}
T_C=\sqrt{\frac{b}{c}}\nu e^{-\frac{1}{4}}.
\end{equation}
Therefore, the above equation is used to calculate the temperature $T_1$ in model\ref{2.2 model}. Although the high temperature approximation is used in the calculation of $T_1$, the results are no different from examining the full potential form given in reference \cite{Hosseini:2024ahe}.

Equations \ref{3.1equation} and \ref{3.2equation} are used to calculate $T_2$. In these equations, the one-loop beta function $\beta_{\lambda_S}$ is needed.
The beta functions of Model \ref{2.2 model} are obtained from the following equations, which use the SARAH package\cite{Staub:2015kfa}:
\begin{align}
& (16\pi^2)\beta_{\lambda_{S}} = 20\lambda_{S}^2 +2\lambda_{SH}^2 +6g_v^4 -12g_v^2 \lambda_{S} ,\nonumber \\
& (16\pi^2)\beta_{\lambda_{SH}}= -\frac{3}{2}g_{1}^2 \lambda_{SH} -\frac{9}{2}g_{2}^2 \lambda_{SH} +12\lambda_{SH}\lambda_{H}+8\lambda_{SH}\lambda_{S} +4\lambda_{SH}^2 + 6\lambda_{SH}\lambda_{t}^2 -6g_v^2 \lambda_{SH}  \nonumber ,\\
& (16\pi^2)\beta_{\lambda_{H}}= +\frac{3}{8}g_{1}^4 +\frac{3}{4}g_{1}^2 g_{2}^2 +\frac{9}{8}g_{2}^4 -3g_{1}^2 \lambda_{H} -9g_{2}^2 \lambda_{H} +24\lambda_{H}^2 +\lambda_{SH}^2 +12\lambda_{H}\lambda_{t}^2 -6\lambda_{t}^4, \nonumber \\
& (16\pi^2)\beta_{g_v}= \frac{1}{3}g_v^3.
\label{RGE}
\end{align}
where $\lambda_t$ is top Yukawa coupling. The parameters $g_1$ and $g_2$ are respectively the $SU(2)_L$ and $U(1)_Y$ SM couplings

Using the results presented above, for Model \ref{2.3 model} and in the high temperature approximation, the critical temperature is obtained from Equation \ref{critical1} where the coefficients $b$ and $c$ are:
\begin{align}
& b =  \frac{1}{64 \pi^{2} \nu^{4}} (M_{H_{1}}^{4}+6M_{W}^{4}+3M_{Z}^{4}+3M_{V}^{4}-12M_{t}^{4}-4M_{\psi}^{4}) , \nonumber \\
& c = \frac{1}{12 \nu^{2}} (M_{H_{1}}^{2}+6M_{W}^{2}+3M_{Z}^{2}+3M_{V}^{2}+6M_{t}^{2}+2M_{\psi}^{2}) . \label{appendix 6}
\end{align}
Of course, the results have also been examined with the full potential of the model, as mentioned in reference \cite{Hosseini:2023qwu}, which does not make much difference.

The beta functions of Model \ref{2.3 model} are:
\begin{align}
& (16\pi^2)\beta_{\lambda_{S}} = 20\lambda_{S}^2 +4g_s^2 \lambda_{S} -12\lambda_{S}g_v^2 -2g_s^4 +2\lambda_{SH}^2+6g_v^4  ,\nonumber \\
& (16\pi^2)\beta_{\lambda_{SH}}= -\frac{9}{10}g_{1}^2 \lambda_{SH} -\frac{45}{10}g_{2}^2 \lambda_{SH} +12\lambda_{SH}\lambda_{H}+8\lambda_{SH}\lambda_{S} -4\lambda_{SH}^2 + 2\lambda_{SH}g_s^2 -6g_v^2 \lambda_{SH}+6\lambda_{t}^4  \nonumber ,\\
& (16\pi^2)\beta_{\lambda_{H}}= +\frac{27}{200}g_{1}^4 +\frac{9}{20}g_{1}^2 g_{2}^2 +\frac{9}{8}g_{2}^4 -\frac{9}{5} g_{1}^2 \lambda_{H} -9g_{2}^2 \lambda_{H} +24\lambda_{H}^2 +\lambda_{SH}^2 +12\lambda_{H}\lambda_{t}^2 -6\lambda_{t}^4, \nonumber \\
& (16\pi^2)\beta_{g_s}=\frac{-3}{2}g_v^2 +2g_s^3, \nonumber \\
& (16\pi^2)\beta_{g_v}= g_v^3.
\label{RGE2}
\end{align}
Using $\beta_{\lambda_S}$ and equations \ref{3.1equation} and \ref{3.2equation}, the temperature $T_2$ can be obtained.

The total thermal effective potential of model \ref{2.4 model} in the high temperature approximation is as follows\cite{Ghorbani:2017jls}:
\begin{equation}
V_{eff}=V_0(h_1 , h_2)+V_{1-loop}(h_1 , h_2;0)+V_{1-loop}(h_1 , h_2;T)\label{appendix 8},
\end{equation}
where $V_0$ is the tree-level potential in eq.\ref{2-19}, $V_{1-loop}(h_1 , h_2;0)$ is the Coleman-Weinberg one-loop correction at zero temperature \cite{Coleman:1973jx} and $V_{1-loop}(h_1 , h_2;T)$ is the one-loop thermal correction\cite{Dolan:1973qd}. In the high-temperature approximation, the one-loop effective
potential takes the following form:
\begin{equation}
V_{1-loop}^{high-T}(h_1 , h_2;T)= (\frac{1}{2}\kappa_{h_1}h_1^2 +\frac{1}{2}\kappa_{h_2}h_2^2)T^2,
\end{equation}
in which we have
\begin{equation}
\kappa_{h_1}=\frac{1}{48}(9g_1^2 +3g_2^2+12g_t^2+24\lambda _H+4\lambda _{SH})\label{4-3},
\end{equation}
\begin{equation}
\kappa_{h_2}=\frac{1}{12}(4\lambda _{SH}+3\lambda _S - g_d^2).
\end{equation}
We have dropped the Colman-Wienberg zero-temperature correction
since at high temperature approximation only the thermal corrections are dominant.
In eq.\ref{4-3} the parameters $g_1$ and $g_2$ are respectively the $SU(2)_L$ and $U(1)_Y$ SM couplings and $g_t$
is the top quark coupling. Full details and the exact equations required to calculate $T_1$ are given in reference \cite{Ghorbani:2017jls}, which we have used.

\bibliography{References}
\bibliographystyle{JHEP}
\end{document}